\documentclass[conference,a4paper]{APSIPA2021}
\usepackage{multirow}
\usepackage{graphicx}
\usepackage{epstopdf}
\usepackage{amsmath,amsfonts}
\usepackage[psamsfonts]{amssymb}
\usepackage{amsxtra}
\usepackage{threeparttable}
\usepackage{subfigure}
\usepackage{booktabs}
\usepackage{bm}

\begin{document}

\title{Noisy-to-Noisy Voice Conversion Framework with Denoising Model}

\author{%
\authorblockN{%
Chao Xie\authorrefmark{1}, Yi-Chiao Wu\authorrefmark{2}, Patrick Lumban Tobing\authorrefmark{2}, Wen-Chin Huang\authorrefmark{1} and Tomoki Toda\authorrefmark{2}
}
\authorblockA{%
\authorrefmark{1}
Graduate School of Information Science, Nagoya University, Nagoya, Japan \\
E-mail: \{xie.chao, wen.chinhuang\}@g.sp.m.is.nagoya-u.ac.jp}
\authorblockA{%
\authorrefmark{2}
Information Technology Center, Nagoya University, Nagoya, Japan\\
E-mail: \{yichiao.wu, patrick.lumbantobing\}@g.sp.m.is.nagoya-u.ac.jp, tomoki@icts.nagoya-u.ac.jp }
}

\maketitle
\thispagestyle{empty}

\begin{abstract}
  In a conventional voice conversion (VC) framework, a VC model is often trained with a clean dataset consisting of speech data carefully recorded and selected by minimizing background interference. However, collecting such a high-quality dataset is expensive and time-consuming. Leveraging crowd-sourced speech data in training is more economical. Moreover, for some real-world VC scenarios such as VC in video and VC-based data augmentation for speech recognition systems, the background sounds themselves are also informative and need to be maintained. In this paper, to explore VC with the flexibility of handling background sounds, we propose a noisy-to-noisy (N2N) VC framework composed of a denoising module and a VC module. With the proposed framework, we can convert the speaker's identity while preserving the background sounds. Both objective and subjective evaluations are conducted, and the results reveal the effectiveness of the proposed framework.
  
\end{abstract}

\section{Introduction}
Voice conversion (VC) is a technique to convert the voice characteristics of a source speaker into that of a target speaker while preserving the linguistic contents.  With the advent of deep learning, VC also enters a new era by dramatically improving the naturalness and similarity of the converted speech.  According to the latest Voice Conversion Challenge (VCC) \cite{zhao2020voice} held in 2020, the state-of-the-art method \cite{liu2020non} shows that the similarity is comparable to natural target speech with slight disparity for naturalness.  

However, in real-world scenarios, we can not always get a large amount of high-quality VC data as it is very costly to collect them.  Although background noise sounds usually interfere with the input speech signal, it would be much appreciated to leverage such mega data to train a VC model in a data-driven technique.  Therefore, it is essential to suppress the background noise to achieve better VC performance.  However, speaking aside from conventional VC, as in VCC, we do not always filter out the background noise obtained from real-world speech signals.  Consider VC usage in a video or a movie; it is essential to only convert the speech segments and preserve the background sounds.  In other cases, such as VC-based speech data augmentation \cite{shahnawazuddin2020voice} for automatic speech recognition (ASR), the background noise is a valuable resource that further improves the robustness of the downstream system.  Therefore, flexibly dealing with the background sounds in VC is more beneficial in general.

The majority of previous research works, such as \cite{miao2020noise,takashima2013noise,takashima2012exemplar} focus on noise-robust VC, in which the background sounds are considered as interference to be discarded.  These works employ the use of noisy input speech and clean target speech.  On the other hand, there has been proposed a text-to-speech method \cite{hsu2019disentangling} that can convert noisy speech while controlling the noise.  To disentangle the speaker identity and the noise attributes, the method augments the clean training set with a copy that mixes with the noise clips but reuses the same transcript and speaker label.  By doing so, two latent variables can be used to represent speaker identity and noise attributes, respectively.  They are modeled by the variational autoencoder (VAE) and introduced to condition the generative process so that both the speaker identity and the background noise are controllable. 

In this paper, we propose a noisy-to-noisy (N2N) VC as a new VC framework, where speaker conversion is achieved while maintaining input background noise without linguistic input/supervision.  The noisy-to-noisy VC signifies that the available dataset for VC training contains only noisy speech signals; therefore, we cannot simply train a noisy-to-clean speech model.  To handle such a noisy speech dataset in VC training, as a first step, we propose to utilize a denoising module for separating speech signal and noise signal, where this denoising module is developed beforehand with publicly available datasets. Then, we propose to employ a conversion network that is developed with the use of the denoised VC dataset. Finally, in the conversion phase, the separated background noise is added back to the converted speech to maintain the input background sounds. We will show that the utilization of the denoising module can enhance the N2N VC system compared to the usage of purely noisy signals in the development of the conversion network. The main contributions of our work are as follows:
\begin{itemize}
\item Unlike previous works focusing on noisy-to-clean VC (noise-robust VC), we aim at noisy-to-noisy VC in our work.  The first "noisy" means only noisy data are available for the VC task.  The second "noisy" indicates that the background sounds are maintained, and we can add the background sounds back or suppress them based on different scenarios. 
\item In this work, we integrate the state-of-the-art denoising model, Deep Complex Convolution Recurrent Network (DCCRN) \cite{hu2020dccrn}, and speaker VC model, vector-quantized variational autoencoder (VQ-VAE), into the proposed N2N VC framework.  We adopt the DCCRN to separate speech and background sounds, and VQ-VAE receives the denoised data as training/conversion input. 
\item To further investigate how the denoising model would influence the downstream VC performance, we insert another famous denoising model Conv-TasNet \cite{luo2019conv} into our framework to compare with DCCRN.
\item We conduct objective and subjective evaluation, and the experimental results indicate that our method achieves an acceptable conversion performance with well-preserved background sounds.
\end{itemize}



\section{Noisy-to-Noisy voice conversion framework}

 \begin{figure}[t]
  \centering
  \includegraphics[width=\linewidth]{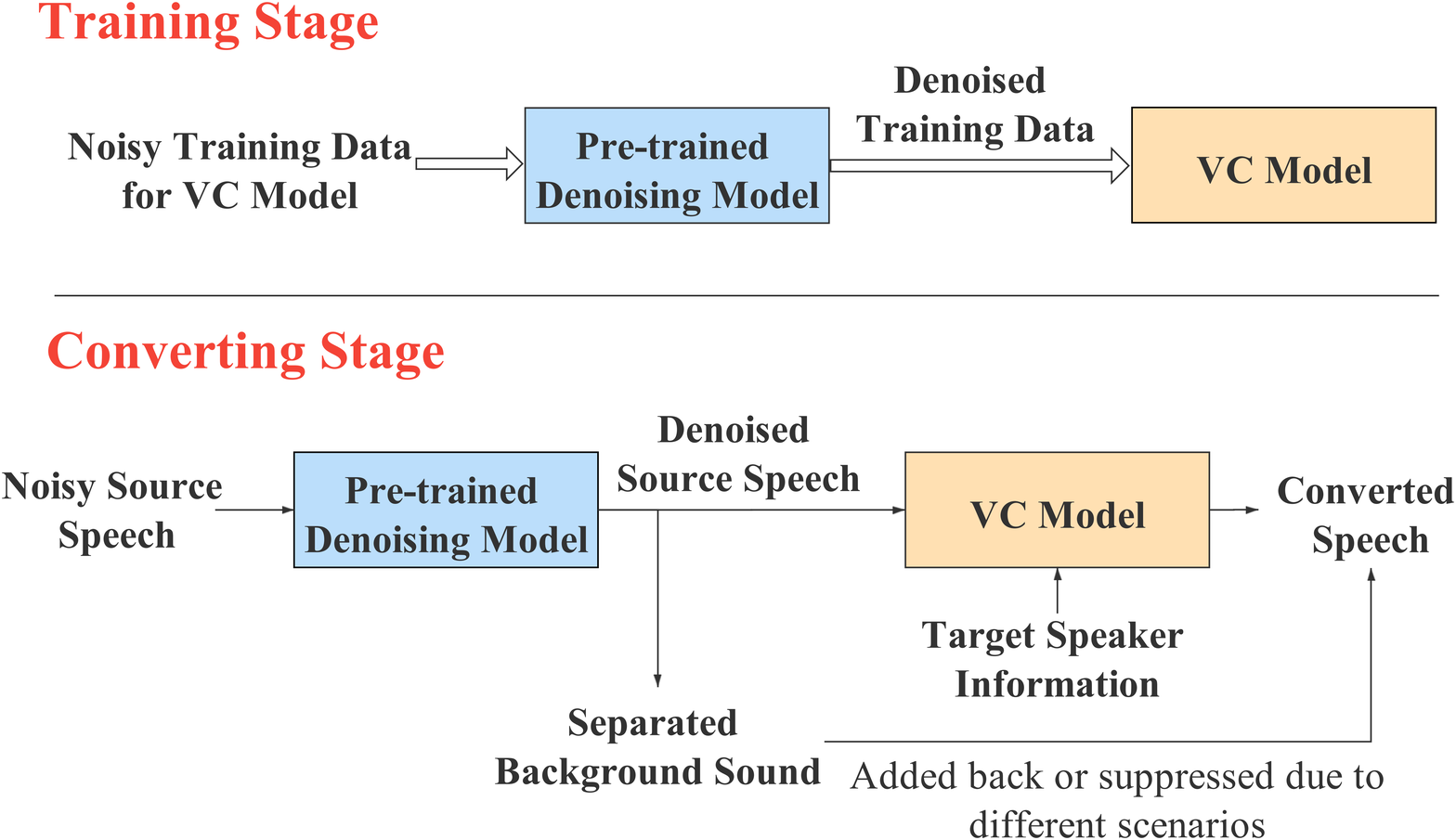}
  \caption{The overall workflow of our proposed N2N VC framework.}
  \label{fig:workflow}
\end{figure}

Our framework is composed of a denoising module and a VC module. Fig.~\ref{fig:workflow} illustrates the overall workflow.   In our framework, the denoising module is pre-trained on the mega dataset to guarantee ideal denoising performance, and it is utilized as a separation model to separate the speech and the background sounds:
\begin{equation}
b(t)= x_{n}(t) - x_{e}(t),
\end{equation}
where \(b(t)\) denotes the estimated background sounds signal in the time-domain. \(x_{n}\) and \(x_{e}\) represent the time-domain noisy speech signal and the estimated speech signal, respectively.  
As mentioned previously, N2N VC is proposed to address the situation that only a noisy VC dataset is available, and the background noise is required to be preserved.  In the training stage, the noisy VC training data pass through the denoising module, and only the denoised data are sent to train the VC module.  In the conversion stage, the noisy source speech is separated by the denoising module, and only the estimated speech signal is delivered to the VC module.  After the conversion, the separated background sounds can be either added back or dropped out, based on individual scenarios.

\section{Framework implementation}

 \begin{figure}[t]
  \centering
  \includegraphics[width=\linewidth]{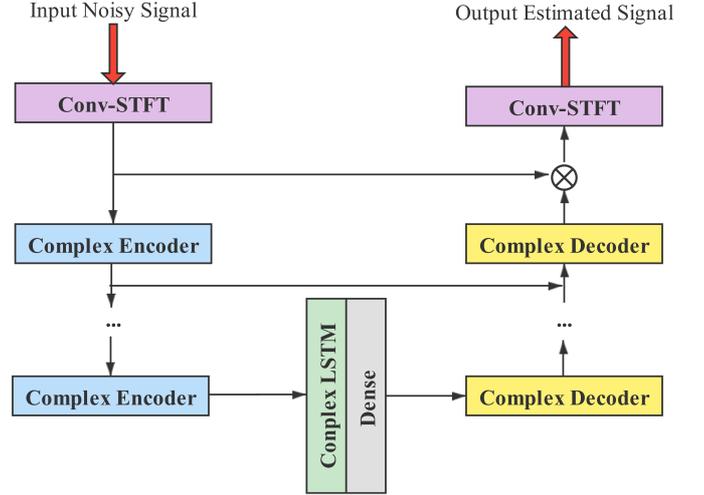}
  \caption{The overall structure of DCCRN.}
  \label{fig:dccrn}
\end{figure}

The motivation of our method comes from the encouraging results of the speech enhancement (SE) domain, wherein the latest Deep Noise Suppression (DNS) Challenge 2020 \cite{reddy2021icassp}, DCCRN \cite{hu2020dccrn} has demonstrated the state-of-the-art performance.  In line with our hypotheses, we expect that a reliably tested denoising module could bring reasonable improvements in developing a conversion network for N2N VC, where we do not have clean signals for the VC dataset.  On the other hand, for the conversion network, we propose to utilize a non-parallel and linguistically unsupervised module based on VQ-VAE, which has been shown to be capable of performing the disentanglement of content and speaker information better compared to conventional variational autoencoder and autoencoder \cite{chorowski2019unsupervised}.

\subsection{Denoising module: DCCRN}

DCCRN is a convolution recurrent network (CRN) based single-channel denoising model.  Fig.~\ref{fig:dccrn} shows the overall structure of DCCRN.  Two-dimensional convolution (Conv2D) blocks are stacked to constitute the encoder/decoder.  Each Conv2D block consists of a convolution/deconvolution layer along with batch normalization and activation function.  The DCCRN has been shown to outperform conventional CRN \cite{tan2019complex} by a large margin thanks to the handling of the problems of complex calculation that are observed in the CRN.   Specifically, complex convolution neural network, complex batch normalization layer, and complex long short-term memory (LSTM) are implemented for encoder/decoder, guaranteeing that the DCCRN can model the correlation between magnitude and phase.  More details can be found in \cite{hu2020dccrn}. 

In our work, since we utilize DCCRN as a separation model, the power of the estimated speech should be matched to the clean target speech.  Hence, the original scale-invariant signal-to-noise ratio (SI-SNR) loss \cite{luo2019conv} is replaced by scale-dependent signal-to-distortion (SD-SDR) loss \cite{le2019sdr}, which is formulated as:
\begin{equation}
\text { SD-SDR }=10 \log _{10}\left(\frac{\|\alpha s\|^{2}}{\|s-\hat{s}\|^{2}}\right),
\end{equation}
where \(s\) and \(\hat{s}\) indicate the target signal and the estimate of the target respectively, and \(\alpha\) denotes an optimal scaling factor defined as:
\begin{equation}
\alpha=\hat{s}^{T} s /\|s\|^{2}.
\end{equation}

 \subsection{VC module: VQ-VAE}
 
 \begin{figure}[t]
  \centering
  \includegraphics[width=\linewidth]{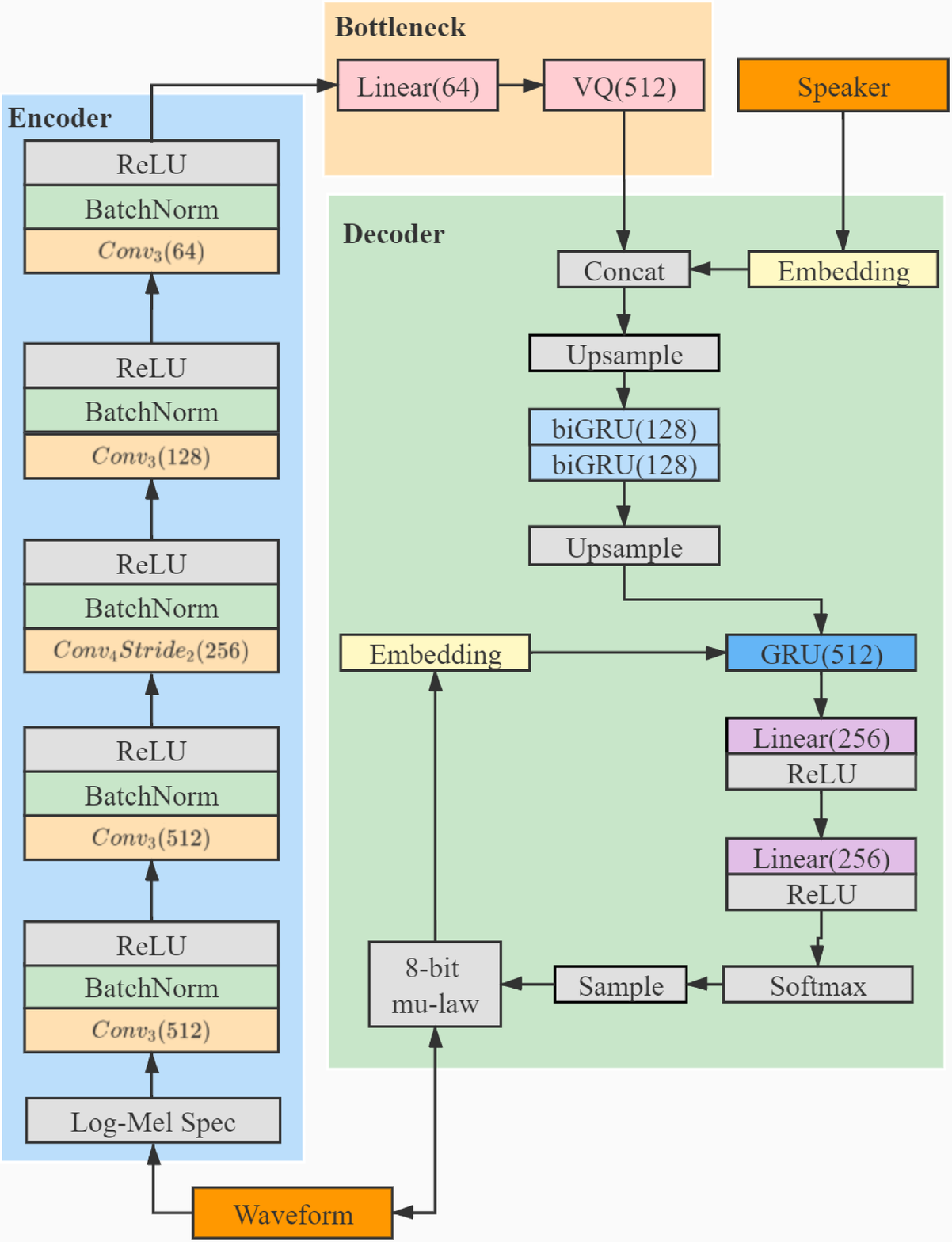}
  \caption{The model structure of the VQ-VAE: An encoder (blue) encodes the log mel-spectrogram into latent representation and passes to the VQ bottleneck (orange). The decoder (green) then reconstructs the waveform from the discrete representation using an autoregressive stream and a speaker embedding as condition. The subscript number of \emph{Conv} represents its convolutional kernel size, and that of \emph{Stride} means the length of stride in the convolutional computation.}
  \label{fig:vqvae}
\end{figure}

In our work, as illustrated in Fig.~\ref{fig:vqvae}, we implement a VQ-VAE-based VC module \cite{van2020vector}, which has three main components: encoder, bottleneck layer, and decoder.

The encoder consists of five one-dimensional convolution (Conv1D) blocks, and each block is composed of a convolution layer along with batch normalization and activation function.  The input log mel-spectrogram sequence $\left\{x_{t}, t=1, \ldots, T\right\}$ is computed as a stream of latent vectors $\left\{z_{j}, j=1, \ldots, N\right\}$ by the encoder and sent to a vector-quantized bottleneck with a 64-dimensional trainable codebook $\left\{e_{i}, i=1, \ldots, 512\right\}$ to discard speaker information.  In the forward pass, the latent vectors of the encoder $\left\{z_{j}, j=1, \ldots, N\right\}$ are mapped into the nearest vectors in the codebook by 
\begin{equation}
k=\arg \min _{i}\left\|z_{j}-e_{i}\right\|^{2},
\end{equation}
and \(z_{j}\) is replaced with \(e_{k}\) as a discrete latent representation $\left\{\hat{z}_{j}, j=1, \ldots, N\right\}$.  The decoder adopts a lightweight recurrent network to reconstruct the waveform based on the embedded speaker identity information and the discrete representation $\left\{\hat{z}_{j}, j=1, \ldots, N\right\}$ from the VQ bottleneck in an autoregressive manner that predicts the current sample based on the past ones.  

In the backward pass, the gradient of the loss through the codebook is approximated via the straight-through estimator \cite{bengio2013estimating}, due to that the \(argmin\) is not differentiable.  The values of the codebokk are updated by exponential moving averages \cite{oord2017neural}. Additionally, a commitment loss \cite{oord2017neural} is introduced to encourage the output vector of the encoder \(z_{j}\) to be close to its selected vector \(e_{k}\) of the codebook.  The VQ-VAE is trained to minimize a sum of two loss terms: the negative log-likelihood of the reconstruction loss and the commitment loss as follows:
\begin{equation}
\mathcal{L}=-\frac{1}{T} \sum_{t=1}^{T} \log p\left(x_{t} \mid \hat{x}_{t}\right)+\beta \frac{1}{N} \sum_{j=1}^{N}\left\|z_{j}-\operatorname{sg}\left(\hat{z}_{j}\right)\right\|^{2},
\end{equation}
where $\left\{\hat{x}_{t}, t=1, \ldots, T\right\}$ is the output sequence of the decoder.  \(\beta\) is the commitment weight and set to 0.25 according to \cite{oord2017neural}, and $\operatorname{sg}(\cdot)$ denotes the stop-gradient operation.

\section{Experimental Evaluations}

We conducted experimental evaluations to investigate the effectiveness of the proposed N2N VC framework. Since we focus on VC application for telecommunication, such as telephone speech conversion or data augmentation for speaker recognition of telephone speech, as one of our target applications of N2N VC, we used 8 kHz sampled speech data in the experimental evaluations.

 \subsection{Dataset}
 
 \begin{table*}[t]
  \caption{The objective evaluation results of DCCRN and Conv-TasNet, higher is better}
  \label{tab:obj-eval-denoising}
  \centering
  \subtable[SI-SDR and SAR on the noisy VCC training dataset]{
  {\begin{tabular}{@{}ccccccccccccc@{}}
\toprule
Model & Eval. Target      & \multicolumn{5}{c}{SI-SDR (dB)}       &  & \multicolumn{5}{c}{SAR (dB)}          \\ \midrule
      &                   & 7 dB   & 11 dB  & 15 dB  & 19 dB  & Avg.  &  & 7 dB   & 11 dB  & 15 dB  & 19 dB  & Avg.  \\ \cmidrule(lr){3-7} \cmidrule(l){9-13} 
\multirow{2}{*}{DCCRN}       & Speech            & 17.86 & 20.17 & 22.73 & 25.19 & \textbf{21.49} &  & 18.55 & 20.82 & 23.32 & 25.76 & \textbf{22.11} \\
                             & Background sounds & 10.48 & 8.55  & 6.87  & 4.86  & \textbf{7.69}  &  & 11.27 & 9.32  & 7.65  & 5.68  & \textbf{8.48}  \\
      &                   &       &       &       &       &       &  &       &       &       &       &       \\
\multirow{2}{*}{Conv-TasNet} & Speech            & 15.39 & 18.01 & 20.46 & 22.77 & 19.16          &  & 16.04 & 18.66 & 21.05 & 23.44 & 19.80          \\
      & Background sounds & 7.45 & 5.64 & 3.42 & 1.06 & 4.39 &  & 8.27 & 6.53 & 4.29 & 1.84 & 5.23 \\ \bottomrule
\end{tabular}}
\label{tab:sub-sisdr}
}
\qquad

\subtable[PESQ and STOI on the noisy VCC training dataset]{        

\begin{tabular}{@{}cccccccccccc@{}}
\toprule
Model       & \multicolumn{5}{c}{PESQ}                       &  & \multicolumn{5}{c}{STOI}                       \\ \midrule
            & 7 dB   & 11 dB  & 15 dB  & 19 dB  & Avg.           &  & 7 dB   & 11 dB  & 15 dB  & 19 dB  & Avg.           \\ \cmidrule(lr){2-6} \cmidrule(l){8-12} 
DCCRN       & 3.20 & 3.41 & 3.57 & 3.72 & \textbf{3.47} &  & 0.96 & 0.97 & 0.98 & 0.99 & \textbf{0.98} \\
            &       &       &       &       &                &  &       &       &       &       &                \\
Conv-TasNet & 2.84 & 3.05 & 3.23 & 3.39 & 3.13          &  & 0.94 & 0.96 & 0.97 & 0.98 & 0.96          \\ \bottomrule
\end{tabular}

       \label{tab:sub-pesq}

}
\end{table*}
 
  \subsubsection{Dataset for denoising model}
 For the training of the denoising model, we used DNS Challenge 2020 dataset \cite{reddy2021icassp}, which is a vast and high-quality dataset for the SE task.  The dataset consisted of two sub-datasets: the clean speech dataset and the noise dataset. 
The clean speech dataset was derived from a dataset of public audiobooks, Librivox \cite{kearns2014librivox}.  The organizers of the DNS Challenge had already cherry-picked the speech files via subjective quality evaluation.  The resulting clean speech dataset had 500 hours of speech from 2,150 speakers in various languages, most of which were in English.  6,000 speech clips were randomly sampled as the validation data. 
 
 The noise dataset was collected from Audioset \cite{gemmeke2017audio} and Freesound \cite{fonseca2017freesound}.  Preprocessed by the organizers, the selected dataset had about 150 audio classes and a total of 65,000 audio clips.  500 clips were randomly picked into the validation set.  We built up the noisy dataset by uniformly sampling a noise clip and adding it to a clean speech.  The SNR levels were also sampled from a uniform distribution between 5 and 20 dB.

  \subsubsection{Dataset for VC model}
  \label{subsubsection:evaldata4vc}
 The dataset for the VC model ought to be unseen for the denoising model.  We chose VCC 2018 dataset \cite{lorenzo2018voice} as the clean speech dataset and PNL 100 Nonspeech Sounds \cite{hu2010tandem} as the noise dataset to simulate the real-world situation.
 
 VCC 2018 dataset is a high-quality and publicly available dataset specialized for VC tasks.  The speech data was recorded by professional US English speakers in a professional studio without significant noise effects.  There were a total number of 972 utterances for training and 420 utterances for evaluation, involving 12 male/female speakers: 8 source speakers denoted as (VCC2SM1, VCC2SM2, VCC2SM3, VCC2SM4, VCC2SF1, VCC2SF2, VCC2SF3, VCC2SF4) and four target speakers denoted as  (VCC2TM1, VCC2TM2, VCC2TF1, VCC2TF2).  Each speaker uttered 81 and 35 sentences for training and evaluation, respectively, resulting in a total of around 13 minutes of audio.

 The PNL 100 Nonspeech sounds consisted of 100 clips and 20 categories of environmental records, such as crowd noise, cry, tooth brushing, and so on.  We uniformly sampled the noise clips to mix with the VCC 2018 train/evaluation dataset at four certain SNR levels: 7 dB, 11 dB, 15 dB, and 19 dB.
 
 For VCC evaluation data, to guarantee that the participants of the subsequent subjective evaluation could concentrate on marking appropriate scores, the number of evaluating utterances was limited to a proper amount.  Four speakers (VCC2SM3, VCC2SM4, VCC2SF3, VCC2SF4) were selected as source speakers, which resembled the non-parallel (SPOKE) task of VCC 2018 \cite{lorenzo2018voice}, and two speakers (VCC2TF2, VCC2TM2) as target speakers.  Hence there were 8 conversion pairs, and each pair had 35 utterances.   Since the evaluation dataset aimed to compare the conversion performance of different systems equally, the same utterances in different speakers shared the same pattern of background sounds.

 \subsection{Model training details}
  \subsubsection{DCCRN training}
  We trained the DCCRN model implemented by Asteroid \cite{pariente2020asteroid}.  The type of DCCRN was "DCCRN-CL."  The window length, hop size, and FFT length were set to 50 ms, 12.5 ms, and 512, respectively.  We observed degradation of the denoising performance with the original settings \cite{hu2020dccrn}, which were for a sampling rate of 16 kHz; hence, we used our own 8 kHz optimized settings.  The batch size was 64.  Adam was used as the optimizer and set the initial learning rate to $1 \cdot 10^{-4}$.  The learning rate would decay 0.5 if the validation loss did not go down within 4 epochs.  Additionally, an early stopping mechanism was introduced to choose an optimized model.  It took about 22 days of training on a single RTX 3080 to get the best model.
  
  \subsubsection{VQ-VAE training}
  We used a PyTorch-based implementation for the VQ-VAE model  \cite{van2020vector}.  Log mel-spectrogram was extracted as the input and 8 bits mu-law decoded waveform as ground-truth.  The window length, the hop size, and the FFT length were set to 20 ms, 5 ms, 1024, respectively. The batch size was 64, and the optimizer was Adam with an initial learning rate of $2 \cdot 10^{-4}$.  The learning rate would be halved after 300k steps.  The total training steps were 600k steps, which cost around two days on a single RTX 3080.

 \subsection{Experimental setup}
 
 To demonstrate the performance of our proposed N2N VC framework, we set two models as our baseline.  The first baseline was a VQ-VAE trained on the clean VCC dataset, denoted as Clean-VC.  The other was the VQ-VAE directly trained on the noisy VCC dataset, denoted as Noisy-VC.  In simpler terms, the Clean-VC and the Noisy-VC respectively represented the upper and lower bound of our framework.
 
 When mixing the noisy VCC dataset, the whole PNL 100 Nonspeech was sampled for both the training and the evaluation set, which indicated that the Noisy-VC had already seen all the patterns of background sounds during training.  Hence, Noisy-VC should have its own optimal performance on the evaluation dataset.  It needs to be emphasized that for our framework, both the noise dataset and speech dataset were unseen for the denoising model.
 
 To further probe into how the denoising model would affect the VC performance in our framework, another well-known denoising model Conv-TasNet \cite{luo2019conv} was selected as a comparison.  Due to the training of the denoising model was very time-consuming, we used the pre-trained Conv-TasNet model provided by Asteroid.  It was trained on the single-speaker enhancement task of the Libri3Mix dataset \cite{cosentino2020librimix}.  To differentiate the use of the denoising models, i.e., DCCRN and Conv-TasNet, we denoted our method as DCCRN-VC and that of ConvTas-VC.  Overall, there were four systems to compare and evaluate:
 \begin{itemize}
\item Clean-VC trained on clean VCC dataset.
\item Noisy-VC trained on noisy VCC dataset.
\item DCCRN-VC trained on DCCRN-denoised noisy VCC dataset.
\item ConvTas-VC trained on ConvTasNet-denoised noisy VCC dataset.
\end{itemize}

   \begin{table}[t]
  \caption{MCD on the clean reference of VCC evaluation dataset, lower is better}
  \label{tab:mcd}
  \centering
  \begin{tabular}{@{}cccccc@{}}
\toprule
Systems      & \multicolumn{5}{c}{MCD (dB)}          \\ \midrule
           & SF-TF & SM-TM & SF-TM & SM-TF & Avg.  \\
Clean-VC   & 7.17 & 7.26 & 7.35 & 7.67 & 7.36 \\
DCCRN-VC   & 7.55 & 7.78 & 7.80 & 8.38 & 7.88 \\
ConvTas-VC & 7.55 & 7.86 & 8.0   & 8.26 & 7.92 \\ \bottomrule
\end{tabular}
\end{table}

\begin{table}[t]
  \caption{Naturalness scores (MOS) with 95\% confidence intervals on noisy VCC evaluation dataset, higher is better}
  \label{tab:mos}
  \centering
  \begin{tabular}{@{}cccc@{}}
\toprule
Systems    & \multicolumn{3}{c}{MOS {[}1, 5{]}} \\ \midrule
           & 7 dB        & 15 dB      & Avg.      \\ \cmidrule(l){2-4} 
Clean-VC   & 3.46 $\pm$ 0.12      & 3.52 $\pm$ 0.11     & 3.49 $\pm$ 0.08     \\
DCCRN-VC   & 3.07 $\pm$ 0.13     & 3.08 $\pm$ 0.12    & 3.08 $\pm$ 0.09     \\
ConvTas-VC & 3.0 $\pm$ 0.13        & 3.14 $\pm$ 0.12    & 3.07 $\pm$ 0.09     \\
Noisy-VC   & 1.99 $\pm$ 0.11     & 2.15 $\pm$ 0.11    & 2.07 $\pm$ 0.08     \\ \bottomrule
\end{tabular}
  
\end{table}

\subsection{Evaluation results}
  \subsubsection{Objective Evaluation}
  \label{subsubsection:obj}
  First, the relative performance of the two denoising models, DCCRN and Conv-TasNet, was assessed with several measurements as follows: scale-invariant signal-to-distortion ratio (SI-SDR) \cite{le2019sdr}, signal-to-artifact ratio (SAR), PESQ \cite{rix2001perceptual}, and STOI \cite{taal2010short}.  In this objective evaluation, the noisy VCC training dataset was used instead of the VCC evaluation dataset, owing to the former that covered the whole of PNL 100 Nonspeech Sounds dataset compared to the latter that was consisted of only 35 different background sound clips at the most.  
  
  The results are demonstrated in Table~\ref{tab:obj-eval-denoising}.  It is evident that DCCRN outperforms Conv-TasNet on all metrics among all SNR levels for both speech and separated background sounds, on which we infer that DCCRN-VC would also provide better performance compared to ConvTas-VC in the following VC evaluations.  We can also observe that as the SNR level increases, the SI-SDR and the SAR of the clean speech increase, while those of the separated background sounds decrease; which is reasonable considering that clean speech could be extracted more easily from a signal with higher signal-to-noise powers (higher SNR) condition rather than from a signal with lower SNR condition, and vice versa for extracting the noise signal.  
  
  To evaluate the performance of our VC model combined with the denoising process, we leveraged clean evaluation reference to assess Clean-VC, DCCRN-VC, and ConvTas-VC via measuring the mel-cepstral distortion (MCD) \cite{kubichek1993mel}.  Table~\ref{tab:mcd} presents the results of MCD on the VCC evaluation dataset.  It can be observed that all three systems achieve better performance for the intra-gender conversions (SF-TF and SM-TM) compared to the cross-gender conversions (SF-TM and SM-TF), where the SM-TF conversion pair is the worst, and the SF-TF conversion pair is the best.  The best average MCD is reasonably achieved by the Clean-VC system with a value of 7.36, and our proposed DCCRN-VC method shows a considerable gap of 7.88 with respect to the Clean-VC. Although DCCRN outperforms Conv-TasNet much in denoising tasks,  the ConvTas-VC with an average MCD value of 7.92 shows only a slightly worse average MCD than the DCCRN-VC.
  
  
  \begin{figure}[t]
\centering
\includegraphics[width=\linewidth]{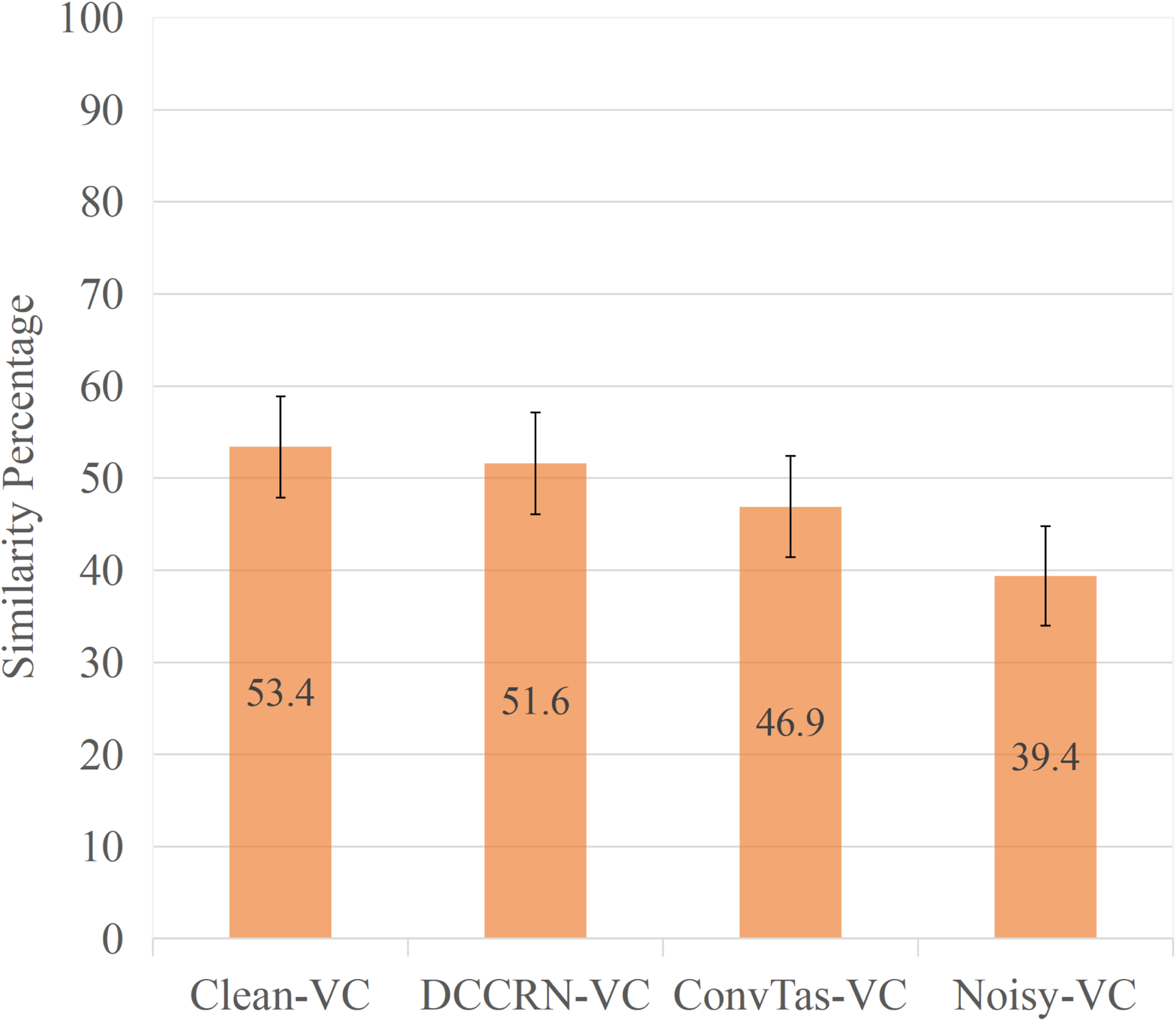}
\caption{Similarity percentage with 95\% confidence intervals on noisy VCC evaluation dataset. The similarity percentage is defined as the added percentage of \emph{Definitely the same} and \emph{Maybe the same}.}
\vspace*{-3pt}
\label{fig:sim}

\end{figure}
  
 Additionally, we also investigated the perceptual quality of the denoised speech and the converted speech.  We observed that the denoised samples of DCCRN sounded clean but with a bit of distortion, while Conv-TasNet remained residual noise throughout the audio.  As for the converted speech, samples from DCCRN-VC and ConvTas-VC sounded with comparable quality.  The distortion by DCCRN leads to further distortion by the VQ-VAE downstream, and the residual noise by Conv-TasNet also degrades the performance of its VQ-VAE.  As we have observed in Table~\ref{tab:mcd}, the resulting MCD values are very close to each other, but as the MCD calculation was performed on only speech segments while the non-speech regions that could still contain the residual noise for the ConvTas-VC were discarded, it is possible that the residual noise in the ConvTas-VC causes adverse effects in a scenario when only clean converted speech needs to be presented.

\subsubsection{Subjective Evaluation} 
  
As our final goal is to achieve high-quality VC while preserving the background sounds, eventually, the overall performance was evaluated through subjective evaluations.  The mean opinion score (MOS) by an opinion test was applied to measure the naturalness of the converted samples.  The participants were asked to give a naturalness score from 1 to 5 (higher is better).  The four systems were evaluated on the same noisy VCC evaluation dataset mentioned in Section~\ref{subsubsection:evaldata4vc}.  To guarantee that the complete subjective evaluation would not take too long so that the participants can submit high-quality answers, we further limited the amount of the evaluation data.  From the evaluation dataset, six utterances were randomly selected for each conversion pair, where three utterances were set for the 7 dB SNR, and the other three were set for the 15 dB SNR.  This resulted in a total number of 204 audio samples: 48 audio samples per system and 12 samples from noisy ground-truth target speech.  As our goal is N2N VC, converted samples from the DCCRN-VC and the ConvTas-VC were superimposed with the respective separated background sounds.  For Clean-VC, we superimposed the original record of background sounds for a fair comparison.  Furthermore, the participants were required to give their scores based on the overall naturalness of both the speech and the background sounds. To assist the participants in making judgment, the category of the superimposed background sounds was given in the evaluation. 

Lastly, we conducted the similarity (SIM) evaluation proposed in \cite{lorenzo2018voice}.  In the SIM test, each of the participants was presented with two audio samples at a time, consisting of a converted speech and a reference speech of the target speaker, and asked to determine whether these samples came from the same speaker. In judging each of the audio pairs, four options were given: \emph{1. Definitely the same; 2. Maybe the same; 3. Maybe different; 4. Definitely different}.  We asked the participants to ignore the quality of the speech and the background sounds and focus on the speaker similarity.  From the evaluation dataset, four utterances were randomly selected for each conversion pair, where two utterances were set for the 7 dB SNR, and the other two were set for the 15 dB SNR.  This resulted in a total of 128 converted samples to be evaluated by each participant and 32 audio samples per system.

The results of the MOS and of the SIM tests are shown in Table~\ref{tab:mos} and Fig~\ref{fig:sim}, respectively, where the SIM score is defined as the sum of the percentages from \emph{Definitely the same} and \emph{Maybe the same} decisions.  Undoubtedly, Clean-VC acquires the best performance with the MOS score of 3.49 and Sim score of 53.4 on average.  Our proposed framework DCCRN-VC reached the MOS score of 3.08 and the SIM score of 51.6 on average, which is far more beyond Noisy-VC that gets 2.07 and 39.4 but still has a margin from Clean-VC.  Thanks to the powerful denoising model DCCRN, our method achieves approximate scores under different SNR levels.  While for Noisy-VC, which is sensitive to noise powers, it reaches better performance under higher SNR level because a higher SNR level means less noise interference.  A similar situation is observed for the ConvTas-VC.

It is worth noting that ConvTas-VC reaches similar naturalness scores on average to DCCRN-VC's.  DCCRN-VC only leads ConvTas-VC with a slight margin, which is consistent with the trend in MCD.  As for ConvTas-VC, due to the inability of Conv-TasNet to completely remove the background noise, the residual noise also exists in the converted samples of the VQ-VAE.  However, after the separated background sounds are superimposed, it is difficult to perceptually notice the interference, which allows ConvTas-VC to obtain a tolerable score in the overall naturalness evaluation.  However, in another scenario, when clean converted speech is required, such residual noise will bring in adverse effects.  It is worthwhile to conduct the subjective evaluation in such a scenario, which would be in our future work.  

As for DCCRN-VC, although we can observe in Table~\ref{tab:obj-eval-denoising} that the performance of the noise removal of the DCCRN is better than that of the Conv-TasNet, as has also been mentioned, the DCCRN introduces some artifacts that are, in turn, propagated to the VQ-VAE VC module.  We believe that this is the reason that the naturalness score of the DCCRN-VC to be in the same range as that of the ConvTas-VC, which would imply that compared to the residual background noise, the unwanted artifacts produced by the denoising module cause more adverse effects to the downstream VC model in terms of audio quality.

\section{Conclusions}

In this paper, we have presented a noisy-to-noisy VC framework that relies on only noisy VC training data and is capable of preserving the background sounds for the converted speech waveform.  Our framework consists of a state-of-the-art denoising model DCCRN and a VC model based on VQ-VAE.  In the training stage, the noisy VC dataset is denoised by the denoising model, on which the VC model is trained. In the conversion stage, the noisy source speech is separated by the denoising model to get the estimated speech signal and background sounds, and the speech signal is sent to the VC model for conversion.  The background sounds can be superimposed or suppressed flexibly according to a specific application.  The experimental results show that our framework outperforms the conventional noisy-to-noisy VC that is directly trained on the noisy VC dataset and achieves acceptable noisy-to-noisy VC performance with room for improvement.  In future work, we aim to bridge the gap between our framework and Clean-VC.

\section*{Acknowledgment}

This work was partly supported by JST CREST Grant Number JPMJCR19A3, Japan.

\bibliographystyle{IEEEtran}
\bibliography{mybib}

\end{document}